\documentclass{elsart}
\usepackage{amssymb}
\usepackage{amsmath}
\usepackage[dvips]{graphicx}

\journal{International Journal of
Engineering Science, Vol. 46  (2008), On line first 2 July 2008 \quad\quad\quad\quad\quad\quad\quad\quad\quad\quad}
\input{tcilatex}

\begin{document}

\begin{frontmatter}

\title{Dynamics of  liquid nanofilms}

\author[UMR6181]{Henri Gouin},
\ead{henri.gouin@univ-cezanne.fr}
\author[UMR6595]{Sergey Gavrilyuk}
\ead{sergey.gavrilyuk@polytech.univ-mrs.fr}

\address[UMR6181]{Universit\'e d'Aix-Marseille \& C.N.R.S.  U.M.R.  6181, \\ Case 322, Av. Escadrille
 Normandie-Niemen, 13397 Marseille Cedex 20 France}
\address[UMR6595]{Universit\'e d'Aix-Marseille \& C.N.R.S.  U.M.R.  6595, IUSTI, Project SMASH,\\ 5 rue E. Fermi, 13453 Marseille Cedex 13 France}

\address {}

\begin{abstract}
The van der Waals forces across a very thin liquid layer
(nanofilm) in contact with a plane solid wall make the  liquid
nonhomogeneous.
The dynamics of such  flat liquid nanofilms is studied in  isothermal case. \\
The Navier-Stokes equations  are unable to describe fluid motions
in very thin films. The notion of surface free energy of a sharp
interface separating  gas and liquid layer  is disqualified.
The concept of disjoining pressure   replaces the model of surface energy.
In the nanofilm  a supplementary free energy    must be considered as a functional of the density.\\
The equation of fluid motions along the nanofilm is obtained
through the Hamilton variational principle by adding, to the
conservative forces, the forces of viscosity in lubrication
approximation. The evolution equation of the film thickness is
deduced and takes into account the variation of the disjoining
pressure along the layer.

\end{abstract}


\end{frontmatter}

\section{Introduction}

The theory of thin liquid layers of a microscopic thickness
is well understood (for a circumstantial bibliography, we may
refer to the review article by Oron et al. \cite{oron}), but the motions of
very thin liquid films  wetting solid substrates are
always object of many debates.
In fact several problems appear: the liquid in very thin layers is
no more incompressible and the equation of motion is  no more Navier-Stokes';
the concept of superficial energy related to a singular surface between gas and liquid layer
has no  more sense.

 Liquids in contact with solids are
submitted to intermolecular forces inferring density gradients at
the walls, making liquids strongly heterogeneous \cite{Israel}.
Often, the fluid inhomogeneity in liquid-vapor interfaces was taken
into account by considering a volume energy depending on space
density derivative \cite {Widom}. However, the van der Waals
square-gradient functional is unable to model repulsive force
contributions and misses the dominant damped oscillatory packing
structure of liquid interlayers near a substrate wall
\cite{chernov1}. Furthermore, the decay lengths are correct only
close to the liquid-vapor critical point where the damped
oscillatory structure is subdominant \cite {Evans1}. Recently, in
mean field theory, weighted density-functional has been used to
explicitly demonstrate the dominance of this structural contribution
in van der Waals thin films and to take into account long-wavelength
capillary-wave fluctuations as in papers that renormalize the
square-gradient functional to include capillary wave fluctuations
\cite {Fisher}. In contrast, fluctuations strongly damp oscillatory
structure and it is  for this reason that van der Waals' original
prediction of a density profile in  \emph{hyperbolic tangent}  form
is so close to simulations and experiments \cite {rowlinson}.
Consequently, depending of  liquids and solids, a great number of
different energy
functionals may be proposed to model a liquid nanofilm in contact with a wall.\\
To compensate the disadvantage of a special functional density to
represent a thin film of liquid, we consider the most general case
of any non-local density free energy functional and deduce a
corresponding \emph{generalized chemical potential}. Then, we
propose an equation of isothermal motions of flat nanofilms in
contact with a plane solid wall. The classical chemical potential
is obtained from the generalized chemical potential in the limit
of homogeneous density. From the classical chemical potential it
is possible to deduce  the disjoining pressure value. The
disjoining pressure exists only when the liquid is   strongly
nonhomogeneous \cite{Nerpin,Derjaguin}. The thin film is driven
along the substrate by the disjoining pressure gradient depending
on the the layer thickness. \\
We must emphasis that the derivation of governing equations is
different from other approaches we can found in the literature.
For example,  in Ref. \cite{oron},  the equation of liquid film dynamics
 is derived in the case of incompressible liquids.
 The disjoining pressure is \emph{ad hoc} added   as a formal additional pressure contribution.

\section{The disjoining pressure for horizontal liquid films}

The following experiment explaining the disjoining pressure
concept  is carefully described in \cite{Derjaguin}. The liquid
bulk of density $\rho_{b}$ was submitted to the pressure $p_{b}$
by means of a very small bubble of radius $R$ attached to a solid
wall (Fig. 1).
\begin{figure}[h]
\begin{center}
\includegraphics[width=5cm]{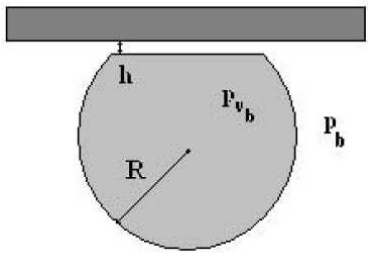}
\end{center}
\caption{\emph{The bubble method of determining the disjoining
pressure isotherms of wetting films} (From \protect\cite{Derjaguin},
page 330).} \label{fig1}
\end{figure}

A thin film of thickness $h$ separates a flat part of the bubble
surface and the solid. The vapor bulk over the layer has a density
$\rho_{v_{b}}$ and a pressure $p_{v_{b}}$. The difference between
the two bulk pressures is
\begin{equation*}
\Pi (h)=\frac{2\sigma}{R}
\end{equation*}
where $\sigma$ is the surface tension coefficient of the bubble; the curve
$\Pi(h)$ obtained by changing the bubble radius is the so-called
disjoining pressure isotherm.

Without repeating the main results of Ref. \cite{Derjaguin} related to
thin liquid films, we enumerate only the properties we use to
describe thin films in contact with a solid wall. \newline We
consider the physical system at a given temperature $\theta $ and
 suppose that the film is thin enough such that the gravity is
neglected across the interlayer.\newline The hydrostatic pressure in
a thin liquid interlayer included between a solid wall and a vapor
bulk differs from the pressure in the contiguous liquid
phase.\newline The forces arising during the thinning of a film of
uniform thickness $h$ produce the  pressure  $\Pi (h)$ which is
equal to the difference between the pressure $p_{v_{b}}$ on the
interfacial surface which is - following the expression given by
Derjaguin - the pressure of the \emph{vapor mother bulk} of density
$\rho _{v_{b}}$, and the pressure $p_{b}$ in the \emph{liquid mother
bulk} of density $\rho _{b}$  from which the interlayer extends
\begin{equation}
\Pi (h)=p_{v_{b}}-p_{b} .  \label{disjoiningpressure}
\end{equation}
At equilibrium, this additional interlayer pressure $\Pi (h)$ is
called the \emph{disjoining pressure}.

The conditions of stability of a thin interlayer essentially
depend on phases between which the film is sandwiched. In case of
a single film in equilibrium with the vapor and a solid substrate,
the stability condition is expressed as a form proposed in Refs.
\cite{degennes}, \cite{degennes2} Chapter 4 :
\begin{equation*}
\frac{\partial \Pi (h)}{{\partial h}}<0.
\end{equation*}
\emph{Let us also notice  an important property}   of a mixture of
fluid and    perfect gas:  the total mixture pressure is the sum
of the partial pressures of components and, at equilibrium, the
partial pressure of the perfect gas is constant through the
liquid-vapor-gas film where the perfect gas is dissolved in the
liquid phase. Calculations and results are identical to those
presented in the following sections: the disjoining pressure of
the mixture is the same as for the fluid without the perfect gas
when a thin liquid film separates liquid and vapor phases
 (see Appendix).

\section{Equation of motion of thin liquid films}
In thin  liquid films the density is strongly inhomogeneous. At
temperature $\theta $ and for a given value $ \mu_0$ of the
chemical potential, the free energy of a liquid film in contact
with a solid is a functional of the density $\rho$
\cite{Fisher1}. The free energy is written in the form:
\begin{equation}
\digamma =\int \int \int_{D}\left( \alpha \,-\rho \mu_0 \right)
dv+\int \int_{\Sigma}\phi \,ds .  \label{free energy
functional}
\end{equation}
In this expression  $D$ denotes the volume of the film and
$\Sigma$  the boundary of $D$. The boundary may  be  a free surface
(as a liquid-vapor interface bordering the film) or a solid-fluid
surface.  The gravity effects are neglected. Moreover,
\begin{equation}
\alpha \equiv\rho \,\varepsilon =K\,[\rho ]  \label{free volume
energy}
\end{equation}
denotes the free volume energy of the fluid and
\begin{equation}
\phi =L\,[\rho ] \label{free surface energy}
\end{equation}
denotes the surface energy. Energies $\alpha$ and $\phi $ are
functionals of the fluid density $\rho $  (for example $\alpha $
and $\phi $ are  both functions of the density and successive gradients
of the density).
\newline The equation of equilibrium is obtained by using the
$G\hat{a}teaux$ \emph{derivative} of $\digamma$:
\begin{equation}
\Omega -\mu_0 =0\quad \rm{or}\quad \rm{grad}\ \Omega = 0,
\label{equilibrium equation}
\end{equation}
where  $\Omega$ is the variational derivative of $\alpha$,
\begin{equation}
\Omega=\frac{\delta \alpha }{\delta \rho}. \label{generalized
chemical potential}
\end{equation}
We call $\Omega$ \emph{generalized chemical potential} of the
fluid. For examples, the following expressions of the volume free
energy may be found in the literature:
\begin{itemize}
\item  The  \emph{grand potential}  in molecular dynamics \cite{Evans1}
\begin{equation}
\alpha =\frac{1}{2}\int \int \int_{D_{0}}\kappa\left( \left|
\mathbf{x-y}\right|
\right) \,\rho \left( \mathbf{x}\right) \rho \left( \mathbf{y}\right) dv_{%
\mathbf{y}}.  \label{grand_potential}
\end{equation}
Then,
\begin{equation*}
\frac{\delta \alpha }{\delta \rho }=\int \int
\int_{D_{0}}\kappa\left( \left| \mathbf{x-y}\right| \right) \,\rho
\left( \mathbf{y}\right) dv_{\mathbf{y}} .
\end{equation*}
\item  The free energy for a  generalized van der Waals theory of capillarity
\cite{Widom,eglit,gouin6}
\begin{equation}
\alpha =\alpha (\rho ,\mathrm{{grad}\,\rho ,\ldots ,{grad}^{n-1}\rho
,\theta ).}  \label{free volume energy}
\end{equation}
 Then,
\begin{equation*}
\Omega=  \frac{\delta \alpha }{\delta \rho }  =
\frac{\partial \alpha }{\partial \rho }-\sum_{p=1}^{n-1}(-1)^{p}\mathrm{div}%
_{p}\left( \frac{\partial \alpha }{\partial {\rm grad}^{p}\,\rho
}\right) ,
\end{equation*}
\end{itemize}
where  $\,\mathrm{{grad}^{p}}\,$ and $\,\mathrm{{div}_{p}}\,$
denotes the gradient and divergence operators iterated $p$ times.

The expression  of the surface free energy functional  is
expressed as an expansion in Ref. \cite{gouin}:
\begin{equation*}
 \phi = -\gamma_1 \rho +\frac{\gamma_2}{2}\rho^2 - \gamma_3 \frac{d\rho}{d n}-2 \gamma_4\,\rho\frac{d\rho}{d n}+\ldots,
\end{equation*}
where $\gamma_i\,   (i =1,\ldots,4)$, are positive constants
associated with intermolecular potentials between fluid and solid,
the fluid density $\rho$ is calculated at the wall and $d /d n$
denotes the normal derivative to the wall.\newline The solution of
the equilibrium problem (\ref{equilibrium equation}) with boundary
conditions coming from the variation of functional (\ref{free energy
functional}) determines an explicit form of the disjoining pressure.
An explicit example of such a calculation based on the van der Waals
square-gradient functional
\begin{equation*}
\alpha =W(\rho )+\frac{\lambda }{2}\left| \nabla \rho \right| ^{2},
\end{equation*}
where $W$ is the volume free energy and $\lambda$ is a constant,
is given in Ref. \cite{gouin}.

In the general case of conservative motions, the governing equations
are obtained through the Hamilton  principle of stationary action
\cite{serrin,berdichevsky}: \newline Let $(t,\mathbf{x} )$ be the
Eulerian coordinates of particles. A particle is labeled by its
position $\mathbf{X}$ in a reference space $D_{0}$. Let $D(t)$ be a
corresponding material volume. The motion of a continuum is a
diffeomorphism from $D_{0}$ onto $D(t)$:
\begin{equation*}
\mathbf{x}=\varphi (t,\mathbf{X}),\quad \mathbf{X}=\chi
(t,\mathbf{x}),\quad \varphi\circ\chi = \textbf{id}.
\end{equation*}
The equation of conservation of mass is
\begin{equation}
\rho \,\mathrm{det}\,F=\rho _{0}(\mathbf{X}),\quad F=\frac{\partial \varphi
(t,\mathbf{X})}{\partial \mathbf{X}},  \label{mass}
\end{equation}
where $\rho _{0}$ is the density defined on $D_{0}$, and $F$ is the
deformation gradient. Let us consider a one-parameter family of virtual motions
\begin{equation*}
\mathbf{X}=\Phi (t,\mathbf{x},\eta )\quad \mathrm{with}\quad \Phi (t,\mathbf{%
x},0)=\chi (t,\mathbf{x})
\end{equation*}
where $\eta $ is a scalar defined in the vicinity of zero. The \emph{%
Lagrangian displacement} associated with this family is \cite
{gouin6,serrin,gavrilyuk}
\begin{equation*}
\delta \mathbf{X}=\frac{\partial \Phi }{\partial \eta }(t,\mathbf{x},0).
\end{equation*}
The displacement vector $\delta \mathbf{X}$ is naturally defined
in Eulerian  coordinates. The Hamilton action is
\begin{equation*}
a=\int_{t_{0}}^{t_{1}}\left( \int \int \int_{D(t)}\Lambda\,dv\right)
dt-\int_{t_{0}}^{t_{1}}\left( \int \int_{\Sigma (t)}\phi \,dS\right)
dt
\end{equation*}
where
\begin{equation*}
\Lambda=\rho \left( \frac{1}{2}\,\mathbf{V}^{2}-\varepsilon \right)
\end{equation*}
is the Lagrangian and $\mathbf{V}$ is the velocity vector .
\newline We consider \emph{virtual displacements}
$(t,\mathbf{x})\in \lbrack t_{0},t_{1}]\times D(t)\ \rightarrow \
\delta \mathbf{X}$ \emph{vanishing in a neighborhood of} $\Sigma
(t)$. From Eq. (\ref{mass}) we get :
\begin{equation*}
\delta \rho =\rho \,\mathrm{div}_{0}\,\delta \mathbf{X}+\frac{1}{\mathrm{det}%
\,F}\ \frac{\partial \rho _{0}}{\partial \mathbf{X}}\ \delta
\mathbf{X}
\end{equation*}
where $\mathrm{div}_{0}$ denotes the divergence in $D_{0}$. The definition
of the Lagrange coordinates\ $\mathbf{X}$ implies
\begin{equation*}
\frac{\partial \mathbf{X}(t,\mathbf{x})}{\partial \mathbf{x}}\,\mathbf{V}+%
\frac{\partial \mathbf{X}(t,\mathbf{x})}{\partial t}=0.
\end{equation*}
Then,
\begin{equation*}
\frac{\partial \delta \mathbf{X}}{\partial \mathbf{x}}\,\mathbf{V}+\frac{%
\partial \mathbf{X}}{\partial \mathbf{x}}\,\delta \mathbf{V}+\frac{\partial
\delta \mathbf{X}}{\partial t}=0,
\end{equation*}
or,
\begin{equation*}
\delta \mathbf{V}=-F\,\frac{d\delta \mathbf{X}}{dt}
\end{equation*}
where
\begin{equation*}
\frac{d}{dt}=\frac{\partial }{\partial t}+\mathbf{V}^{T}\rm{grad }
\end{equation*}
denotes the material time derivative and    $^T$   the transposition. If we denote by $N=\displaystyle\frac{1%
}{2}\,\mathbf{V}^{2}-\Omega $, we obtain
\begin{equation*}
\delta a=\int_{t_{0}}^{t_{1}}\left[ \int \int \int_{D(t)}\left\{ N \,\delta \rho +\rho \,\frac{%
d\left( \mathbf{V}^{T}F\right) }{dt}\ \delta \mathbf{X}\right\} dv\right]
\,dt
\end{equation*}
\begin{equation*}
=\int_{t_{0}}^{t_{1}}\left[\int \int \int_{D_{o}}\rho _{0}\left\{
\frac{d\left( \mathbf{V}^{T}F\right)
}{dt}-\mathrm{{grad}_{0}^{\emph{T}}\,\emph{N}}\right\} \delta
\mathbf{X}\,dv_0\right]dt
\end{equation*}

where $\mathrm{{grad}_{0}}$ is the gradient in $D_{0}$.
 \begin{itemize}
 \item  \emph{The Hamilton  principle of least action }states
\begin{equation*}
\forall\ (t,\mathbf{x})\in \lbrack t_{0},t_{1}]\times D(t)\ \longrightarrow \delta \mathbf{X}, \rm{(vanishing\ in\ the\ neighborghood\ of}\ \Sigma(t)),
\end{equation*}
\begin{equation*}
\delta a\equiv\delta \int_{t_{0}}^{t_{1}}\left( \int \int
\int_{D(t)}\Lambda\,dv\right) dt=0 .
\end{equation*}
 \end{itemize}

Consequently, the equation of motion is
\begin{equation*}
\frac{d\left( \mathbf{V}^{T}F\right)
}{dt}=\mathrm{{grad}_{0}^{T}\,\emph{N}.}
\end{equation*}
Let us note that
\begin{equation*}
\frac{d\left( \mathbf{V}^{T}F\right) }{dt}=\left( \mathbf{\Gamma }^{T}+%
\mathbf{V}^{T}\frac{\partial \mathbf{V}}{\partial \mathbf{x}}\right) \,F
\end{equation*}
where $\mathbf{\Gamma }$ is the acceleration vector. For a
conservative isothermal motion, the equation of motion is
\begin{equation}
\mathbf{\Gamma }+ \rm{grad}\, \Omega  =0.  \label{equation of motion1}
\end{equation}
In nonconservative cases, to take into account dissipative
effects, we  simply introduce a symmetric viscous stress tensor
$\mathbf{\ \sigma }_{v}$ in the second member of Eq.
(\ref{equation of motion1}); the equation of motion becomes:
\begin{equation}
\mathbf{\Gamma }+\rm{{grad\, } \Omega =\frac{1}{\rho }\,(div\,\mathbf{%
\ \sigma }_{v})} .  \label{equation of motion2}
\end{equation}
We consider a viscous stress tensor in the form
\begin{equation*}
\mathbf{\ \sigma }_{v}=\kappa _{1}(\text{ tr }{D})\,\mathbf{1}+2\,\kappa
_{2}\;D
\end{equation*}
where
\begin{equation*}
D=\frac{1}{2}\left( \frac{\partial \mathbf{V}}{\partial \mathbf{x}}+\left(
\frac{\partial \mathbf{V}}{\partial \mathbf{x}}\right) ^{T}\right)
\end{equation*}
is the rate of deformation tensor
 and $\kappa _{1}$, $%
\kappa _{2}$ are the viscosity coefficients.\newline Let us notice
the important property: the gradient of the \emph{generalized
chemical potential} $\Omega$ of thin layer is the only driving
force of the liquid.

Let  us remark that Refs. \cite{oron,Nerpin} considered the liquid motion
governed by the Navier-Stokes equations.  The thermodynamic pressure
was simply replaced with the disjoining pressure and the sign of the
pressure gradient was changed in its opposite. An adherence
condition was prescribed at the free liquid-vapor boundary layer.

\section{Motions along a liquid nanolayer}

We consider a horizontal plane liquid interlayer contiguous to its
vapor bulk and in contact with a plane solid wall. We use an
orthogonal system of coordinates such that the $z$-axis is
perpendicular to the solid surface. The liquid film thickness is
denoted by $h$.\newline When the liquid layer thickness $h$ is small
with respect to longitudinal dimensions $L$ along the wall $\left(
\epsilon =\dfrac{h}{L}\ll 1\right) $ and the inertia effects are
negligeable, it is possible to simplify Eqs (\ref{equation of
motion2}) (approximation of
lubrication  \cite{batchelor}). We denote the velocity by $\mathbf{V}%
=(u,v,w)^{T}$ where $(u,v)$ are the tangential components. Due to
the conservation of mass, ${\partial u}/{\partial x}$, ${\partial
v}/{\partial y}$ and ${\partial w}/{\partial z}$  are of the same
order. The main terms associated with second derivatives of liquid
velocity components correspond to ${\partial ^{2}u}/{\partial
z^{2}}$ and ${\partial ^{2}v}/{\ \partial z^{2}}$.\newline
 In the viscous stress tensor  $\mathbf{\
\sigma }_{v}=\kappa _{1}(\text{div}{\mathbf{V}})\,\mathbf{1}+2\,\kappa _{2}\;{D}$,   the order of magnitude of partial derivatives implies
\begin{equation*}
\mathbf{\
\sigma }_{v}\simeq 2\,\kappa _{2}\;{D} .
\end{equation*}
As in \cite{Rocard}, we assume that the kinematic viscosity
coefficient $\nu =\kappa _{2}/\rho $ depends only on the
temperature. Consequently,
\begin{equation*}
 ({1}/{\rho })\text{ div }\mathbf{\sigma }_{v} \simeq 2\nu \, \left\{ \,
\text{div } D \,+\,D\text{ grad [\thinspace Ln}\,(\rho)]\, \right\}.
\end{equation*}
In the liquid nanolayer, the liquid density variation $|(\rho-\rho_b)/\rho_b|\ll 1$
and consequently $D$ grad\{Ln ($\rho$)\} is negligible with respect to div $D$. Then,
\begin{equation*}
 ({1}/{\rho })\text{ div }\mathbf{\sigma }_{v} \simeq 2\,\nu \
\text{div } D .
\end{equation*}
Hence, in the lubrication approximation the liquid nanolayer motion
verifies
\begin{equation}
\text{grad}\,\Omega =\nu \,\Delta {\mathbf{V}}\quad \mathrm{with}\quad
\Delta {\mathbf{V}}\simeq
\begin{bmatrix}
\displaystyle\;\frac{\partial ^{2}u}{\partial z^{2}},\displaystyle\;\frac{%
\partial ^{2}v}{\partial z^{2}},0
\end{bmatrix}.
\label{motion}
\end{equation}

We denote by $\mu(\rho )$ the phase chemical potential obtained
from $\Omega$ for homogeneous densities {$\rho _{l}$ and $\rho
_{v}$ of a plane liquid-vapor interface without solid wall and
having the same value  in the liquid and vapor bulks (e.g.
$\mu\left( \rho _{l}\right) =\mu(\rho _{v})=0$). In presence of a
solid wall, the generalized chemical potential $\Omega $ takes a
bulk value $\mu(\rho _{b})$ $=\mu(\rho _{v_{b}})= \mu_0$ (see Eq.
(\ref{equilibrium equation})).
 The value $%
\rho _{v_{b}}$ is the value of the vapor density at $z\rightarrow
+\infty ,$ and $\rho _{b}$ is a value of the \emph{mother}
homogeneous fluid bulk  in equilibrium with the vapor
\cite{Derjaguin}. The $z$-component of {Eq. (\ref{motion})
yields
\begin{equation*}
\frac{\partial \Omega [\rho ]}{\partial z}=0\quad {\rm and\ consequently}\quad
\Omega [\rho]={\mu} (\rho _{b}).
\end{equation*}
To each value of $\rho _{b}$ (different from  the liquid bulk
density value $\rho _{l}$ of the plane interface at equilibrium)
is associated a liquid nanolayer thickness $h$. Then, the
functional $\Omega [\rho]$ is a function of $h$: $\ \Omega
[\rho]=\tilde{\mu}(h)$ with $\tilde{\mu}(h)=\mu(\rho _{b}(h))$.
The tangential components of Eq. (\ref{motion}) yield for the
velocity} $\mathbf{u=}(u,v)^{T}$ :
\begin{equation}
\rm{grad}\, \mu(\rho _{b})=\nu \frac{\partial ^{2}\mathbf{u}}{\partial z^{2}}%
\ \ \ \ \Longleftrightarrow \ \ \ \frac{\partial \mu}{\partial \rho _{b}%
}\ \rm{grad}\, \rho _{b}=\ \nu \frac{\partial
^{2}\mathbf{u}}{\partial z^{2}}, \label{viscosity}
\end{equation}
where \,grad\, denotes   the two-dimensional gradient with respect
to $(x, y)$.

Let us note that a liquid can slip on a solid wall only at a
molecular level \cite{Chuarev}. The longitudinal sizes of solid
walls are several orders of magnitude higher than slipping distances
and consequently,  in a macroscopic mechanical model along the wall,
the kinematic condition at solid walls is the adherence condition:
\begin{equation*}
\left. {\mathbf{u}}\right| _{z=0}{=0}.
\end{equation*}
From the continuity of fluid tangential stresses through the
liquid-vapor interface of molecular size bordering the liquid
nanolayer and assuming that vapor viscosity stresses are negligible,
we get
\begin{equation*}
\;\left. \frac{\partial \mathbf{u}}{\partial z}\right| _{z=h}=0.
\end{equation*}
Consequently, Eq. (\ref{viscosity}) implies
\begin{equation*}
\nu \,\mathbf{u}= \, \left( \frac{1}{%
2}\,z^{2}-h\,z\right) \, \rm{grad}\, \mu\left(\rho _{b}\right).
\end{equation*}
The mean spatial velocity $\overline{\mathbf{u}}$ of the liquid in
the nanolayer is
\begin{equation*}
{ \overline{\mathbf{u}}=\frac{1}{h}\int_{o}^{h}\mathbf{u}\ dz},
\end{equation*}
and consequently,
\begin{equation*}
\nu \ {\mathbf{\overline{u}}}=-\frac{h^{2}}{3}\ \rm{grad}\,
\mu(\rho _{b}).
\end{equation*}
The chemical potential  is homogeneous in the bulk; then,
\begin{equation*}
d\mu(\rho _{b})=\frac{dp\left(\rho _{b}\right) }{\rho _{b}}\,,
\end{equation*}
where $p\left( \rho _{b}\right) $ is the pressure corresponding to
the bulk density $\rho _{b}$. Hence,
\begin{equation*}
\rm{grad}\,   \mu(\rho _{b})=\frac{\partial \mu}{\partial \rho _{b}}\,%
\frac{\partial \rho _{b}}{\partial h}\,\rm{grad}\, h\equiv \frac{1}{\rho _{b}}\,%
\frac{\partial \emph{p}(\rho _{b})}{\partial \rho _{b}}\,\frac{\partial \rho _{b}}{%
\partial h}\,\rm{grad}\, h\,.
\end{equation*}
The pressure $p_{v_{b}}=p\left( \rho _{v_{b}}\right) $ in the vapor
bulk is constant and the disjoining pressure is
\begin{equation*}
{\Pi (h)=p\left( \rho _{v_{b}}\right) -p(\rho _{b}).}
\end{equation*}
Then,
\begin{equation*}
\rm{grad}\,   \mu(\rho _{b})=-\frac{1}{\rho _{b}}\,\rm{grad}\, \Pi
(h)\, ,
\end{equation*}
and we obtain
\begin{equation}
\chi _{b}\ {\mathbf{\overline{u}}}=\frac{h^{2}}{3}\ \text{grad}\ \Pi
(h), \label{variation potentiel chimique}
\end{equation}
where $\chi _{b}=\rho _{b}\nu $ denotes the  dynamic viscosity of
the liquid. The mean liquid velocity is driven by the variation of
the disjoining pressure along the solid wall and the film square
thickness.

Eq. (\ref{variation potentiel chimique}) is completely different
from the classical hydrodynamics of films:

Indeed, for a thin liquid film, the Darcy law is  $
 {\mathbf{\overline{u}}} = - K(h)\, {\rm grad}\, p$, \ where $p$ is the
 liquid pressure and $K(h)$ is the permeability coefficient. In
Eq. (\ref{variation potentiel chimique}), the sign is opposite and
the liquid pressure is replaced with the disjoining pressure.
The mass equation averaged over the liquid depth is :
\begin{equation*}
\frac{\partial}{\partial t}\left(\int_0^h \rho\,dz\right)+ \rm{div}\left(\int_0^h \rho\,\mathbf{u}\,dz\right)= 0.
\end{equation*}
Since the variation of density is small, the equation for the free
surface is
\begin{equation}
\frac{dh}{dt}+h\ \mathrm{div}{\ \mathbf{\overline{u}}}=0.  \label{h-equation}
\end{equation}
Replacing (\ref{variation potentiel chimique}) into
(\ref{h-equation}) we finally get
\begin{equation}
\frac{\partial h}{\partial t}+\mathrm{div}\left(
\frac{h^{3}}{3\,\chi _{b}}\, \rm{grad}\, \Pi (h)\right) =0.
\label{h-evolution equation}
\end{equation}
Eq. (\ref{h-evolution equation}) is a non-linear parabolic
equation with  a \emph{good sign,} if $ {\
\displaystyle\frac{\partial \Pi (h)}{\partial h}<0 } $. This
result is in accordance with a physical stability  argumentation
about thin liquid layers obtained by \cite{Derjaguin,degennes2}.

\section{Conclusion}
In this paper we use the variational principle of stationary action
to obtain the equation of motion of a liquid in a very thin plane
layer. The result is obtained whatever is the functional of free
energy in the liquid and the disjoining pressure gradient is the
driving force of the film motion. This result is obtained in a
direct way without any phenomenological assumption as it is done in
the literature where the disjoining pressure of a thin film is added
formally in the equation of motion of incompressible liquids. The
disjoining pressure expresses the excess free energy of the layer
with respect with the superficial energy of a fluid interface
(superficial tension)\cite{Lifshitz}. Consequently, in the equation
of motion, the superficial tension has   no raison to be taken into
account. This result can be different if some curvature appears in
the film as in the case of a micro droplet. We have also obtained
the evolution equation for very thin isothermal flat liquid films.
The thin film is driven along the substrate by the disjoining
pressure gradient associated with the layer thickness. The equation
of the film thickness evolution is a nonlinear parabolic equation
having a good sign when the thin liquid layer is stable, the
instability being probably associated
with the rupture of the film along the substrate. \\
In appendix, we will see that the results are unchanged when the
layer is constituted with a mixture of fluid and gas.

\section{Appendix: Extension of results for  a fluid mixture}
Let us consider a mixture of the perfect  gas and the non-homogeneous
liquid.  A mixture of the perfect gas and the saturated vapor  of the
liquid rises above the thin layer.   To model the mixture at
temperature $\theta $, we consider  a volume free energy in the
form \cite{gouin2}:
\begin{equation*}
\rho \varepsilon =\alpha + \beta ,
\end{equation*}
where $\alpha = K[\rho_f]$ is defined as in Eq. (\ref{free volume
energy}) and $\beta = H(\rho_g)$ is the free volume energy of the
perfect gas. The mixture density is $\rho =\rho _{f}+\rho _{g}$
where $\rho_f$ is the fluid density and $\rho_g$ is the perfect gas
density.

In conservative case, without diffusion between constituents and
neglecting the gravity forces, the equations of motion  are \cite
{gouin3} :
\begin{equation}
\left\{
\begin{array}{c}
\ \ \ \  \mathbf{\Gamma_\emph{f}}+\text{grad}\ \Omega[\rho_f] =0 ,\label{melange} \\
\ \ \ \ \mathbf{\Gamma_\emph{g}}+\text{grad}\  \omega(\rho_g) =0 ,
\end{array}
\right.
\end{equation}
where $\mathbf{\Gamma_\emph{i}} \  (i =f,g) $ are the accelerations
of components, $ \Omega[\rho _{f}]$ is the \emph{generalized
chemical potential} of the fluid (Eq. (\ref{variation potentiel
chimique})) and $\omega(\rho _{g})$ is chemical potential of the
perfect gas.
\newline
At equilibrium, the thin layer is invariant along the solid wall and
Eq. (\ref{melange}$_2$)  yields $\rho_g =\rho_{g0}$ where
$\rho_{g0}$  is a constant in the mixture. The total pressure in the
mixture bulks is the sum of the partial pressures of constituents:
$p = p_f + p_g$. The partial pressure of the perfect gas is constant
in  the domain occupied by the mixture and the disjoining pressure
is:
\begin{equation*}
 \Pi =  (p_{fv}-p_g)-(p_{fb}-p_g)=p_{fv}-p_{fb}.
\end{equation*}
Hence, the disjoining pressure can be reduced to the disjoining
pressure for a single inhomogeneous fluid. If we take into account
the liquid viscosity and neglect the perfect gas viscosity,
Eq.(\ref{melange}$_{1}$) becomes:
\begin{equation*}
\mathbf{\Gamma_\emph{f}}+\text{grad}\ \Omega [\rho _{f}] =\nu _{f}\,
\Delta \mathbf{V},
\end{equation*}
where $\nu_f$ is the fluid kinematic viscosity. This result can be extended to  multi-component cases.

\end{document}